# Steven Weinberg (1933–2021)

This is a homage to the memory of Prof. Steven Weinberg who passed away on 23 July 2021. The world of theoretical particle physics is the poorer because of it and feels a great sense of loss. Both of the authors are among the large worldwide community of physicists and astrophysicists who have spent their lifetime working in the areas pioneered by him. He dominated the world of theoretical particle physics from the sixties to eighties, with an amazing breadth of the subjects which he contributed to and an equally amazing depth of these contributions. The only other theorist with comparable breadth of interests in the period was his colleague at the University of Texas at Austin (UT), the Indian physicist: late E. C. G. Sudarshan. Weinberg's idea of the unified description of the electromagnetic and weak interactions (the EW unification), announced in a three-page paper titled 'A model of leptons', is at the centre of the Standard Model (SM) of Particle Physics, a cornerstone of our understanding of fundamental laws of Nature that govern the world of elementary particles. It is in fact now an integral part of the journey of science and scientists to arrive at a basic understanding of the Cosmos and the Universe. In fact, the SM has driven experimental particle physics for six decades. However, it needs to be noted that his work on the EW theories for which he was awarded the Nobel prize (along with Sheldon Glashow and Abdus Salam) in 1979 and which became well known in the world of non-scientists as well since the discovery of the Higgs boson at the Large Hadron Collider (LHC), is just one part of the vast vista of his contributions to particle physics and cosmology.

This giant of 20th century theoretical physics, was born in New York city in 1933 to immigrant parents. He was the first one from his family to go to college. He praised the Bronx High School for Science which he attended for the atmosphere if not for the subjects they taught. He described his time there as cool, where the co-students looked up to the smart students who learnt many topics in science and mathematics, not taught in the school, on their own. (Incidentally 7 pupils from this school went on to become Nobel Laurates, Steven Weinberg, Sheldon Glashow, Melvin Schwarz, Roy Glauber and H. David Politzer to name a few.) According to him, he simply loved the ability to calculate things. He was thrilled when he could calculate the catenary shape of cables in a suspension bridge, though he wanted to move forward from just having fun in solving problems where the solutions were known.

After finishing school, he obtained his bachelor's degree at Cornell University in 1954. After spending one year at Copenhagen at the Niels Bohr Institute he joined Princeton University. It is this one year in Copenhagen where he began doing independent research on a small problem

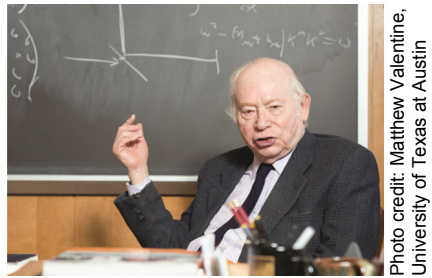
Photo credit: Matthew Valentine, University of Texas at Austin

suggested by Gunnar Källén. It is this experience that was behind his advice to young people, given much later at the graduation address at McGill University: 'You don't have to know everything because I didn't when I got my PhD'. He advised young people to learn things on the job as it were, as one goes along working on the subject!

After finishing in his Ph.D. in 1957 under the supervision of Sam Treiman at Princeton University, he moved as a postdoctoral research fellow, first to Columbia University and then to the University of California, Berkeley in 1959. He stayed there till 1966 as faculty during which period he worked on a broad canvas of Quantum Field Theory, its version called Effective Field theory (EFT), and Gravitation. In 1966 he moved to Boston, holding positions at MIT and later at Harvard, where his wife was studying for her law degree. He says in an interview 'a lot of my thinking was done while sitting on the park benches and watching my daughter play'. Around 1979 Weinberg and Mrs Louise Weinberg (nee Goldwasser), professor of law at Harvard started exploring to move to another University. He began to teach at UT as early as Spring of 1980 and formally became Josey Regental Professor in the Departments of Physics as well as Astronomy by 1982. In his arrival at UT, he added even more to the very high profile it already had, with stalwarts Ilya Prigogine, John A. Wheeler, Bryce de Witt and E. C. G. Sudarshan being present there. L. Weinberg became a professor in the UT School of Law, a position which she holds till date.

The late fifties and early sixties was a time of great confusion in the world of theoretical particle physics. Experimental particle physics was way ahead of the theories, producing information on various aspects of the world of hadrons (n, p, pions, kaons, …) involving the strong and the weak interactions. In the case of strong interactions there were many models which could be constructed to describe the same but none of them had any predictive power due to the large strength of the interaction which made perturbative calculations impossible. For the weak interactions one had a generalization of Fermi's theory in the form of 'V–A' theory of E. C. G. Sudarshan and R. E. Marshak, also proposed by R. P. Feynman and M. Gell-Mann, which by itself had provided a substantial order in the world of weak interactions. But unfortunately, this theory stopped making any sense at high energies. Weinberg made pioneering, all important contributions which brought order in both these areas. The one in the context of weak interactions was the above-mentioned EW unification and the construction of 'A model of leptons'. This paper written in 1967, when he was a visiting professor at MIT, is *the* highest cited paper in Particle Physics ever. The work in the context of strong interactions provided a way to expand the amplitudes in inverse power of energy rather than powers of the strong coupling. This goes by the name of Effective Field Theories (EFT). In fact, in view of the lack of any hint of a new particle or interactions at the LHC so far, the EFT methods have emerged once again as frontline tools to probe the physics of interactions and/or particles beyond those in the SM (BSM physics), at the LHC and otherwise. It has also led to formalism of the Soft-collinear Effective Theories (SCET) which is used to compute quantities of interest in theories with widely different scales. In fact, it finds applications in Condensed Matter Physics as well.

Weinberg along with Goldstone and Abdus Salam had been worrying about the famous phenomenon of spontaneous symmetry breaking. But as he himself said in a talk, he was looking at it with a view to address the issue of a theoretical description of the observed features of strong interactions. Somewhere along the line hit the realization that this could actually hold the promise of solving the high energy problem of weak interactions. The clarity it brought to our understanding of the electromagnetic and weak interactions can only be compared with the one provided to the subject of electromagnetic interactions by the formulation of Maxwell's theory giving a unified description of electricity and magnetism. That was confirmed by the agreement of the measurement of the velocity of light with the one predicted in terms of the dielectric constant and magnetic permeability of vacuum. Similarly, 'A Model of Leptons' included the prediction of the masses of the charged W and the postulated, weak neutral Z boson, as well as their interactions with the leptons (electrons, muons and associated neutrinos) in terms of a single parameter denoting the EW mixing. (For quite some time it was referred to in literature as the Weinberg angle!) Hence the 1979 Nobel prize could be given for EW unification, based on the discovery of neutral weak currents at CERN in 1973, even when the W and the Z had not been experimentally discovered.

This formulation posited the existence of a new, spin-0 boson which we now know as the Higgs boson. It took Weinberg's genius to realize that the same Higgs boson could help us understand how the spin ½ fermions (in the case of his model the charged leptons) could also be massive still maintaining symmetry properties that gave the theory its good high energy behaviour. This model of the 1967 paper extended to include quarks and their strong interactions, is now called the 'Standard Model'. Establishing and testing it has kept particle physicists busy for six decades. Finding a disagreement with the SM predictions has been now the holy grail of experimental high energy physicists for a few of these decades.

What did the virtuoso physicist have to say about this epochal piece of work? At the time itself he had said 'it has too many arbitrary features for [its] predictions to be taken very seriously'. In an interview to CERN courier on the occasion of his winning the 'breakthrough prize in fundamental physics' he said, 'But it was rather untypical of me. My style is usually not to propose specific models that will lead to specific experimental predictions, but rather to interpret in a broad way what is going on and make very general remarks, like the development of the point of view associated with Effective Field Theory'. The same EFT approach, in fact provides a way to understand why the observed neutrino masses can be 'naturally' so much smaller than masses of all the other quarks and leptons. In fact, the lepton and quark masses, which in the SM are understood in terms of the arbitrary parameters that he so abhors in the quote above, had been a matter of concern to him till the end. He wrote an interesting paper on 'Model for quark and lepton masses' as late as January 2020.

As is indicated in the quote in the context of the SM that is given above, Weinberg's work was characterized by a zeal to bring intellectual clarity and conciseness to the frameworks being deployed, in contrast to the utilitarian approach of the physics community to tolerate redundancies and ambiguities so long as the toolbox of methods worked. Quantum electrodynamics (QED) had enjoyed tremendous success in the hands of Schwinger, Feynman and had been duly rationalized as an elegant framework of perturbation theory by Dyson by 1950. Aside from the problems with the high energy behaviour of quantum field theory (QFT) there was much confusion about incorporating particles of ever higher spin which were being discovered at the accelerator experiments. In a series of papers in the early 1960s Weinberg developed what may be considered a 'constructive approach' to QFT and developed a systematic approach to handle particles of any spin.

This streamlined exposition of QFT was taught by him as a series of graduate courses, and generations of graduate students learnt from the books which he wrote on the subject. One of us (UY) had the opportunity to take them over a period of two years in mid-1980s. A set of cryptic handouts with essential calculations appeared periodically, incomprehensible to anyone who did not know the context. To a mild suggestion to include at least the most critical pointers within the notes, he laughed and said, 'for that you have to come to class'. He was an exceptional teacher and continued to teach at the UT right till the end. He used to say, 'As is natural for an academic, when I want to learn about something, I volunteer to teach a course on the subject.'

The decade of 1970s was marked by most fantastic theoretical developments in elementary particle physics and the idea of unification which started from Faraday and Maxwell through EW unification led to the possibility of Grand Unification, that all interactions may be unified. Weinberg with Helen Quinn and Howard Georgi, predicted the rate at which protons should decay (with a life time of $10^{30}$ years), suggesting a specific experimental test of the hypothesis. This inspired a search for such signals in the Kolar Gold Fields experiment of TIFR and also motivated the construction of the Kamiokande experiment in Japan. While no signals of proton decay have emerged till date, pushing the limits on the lifetime of the proton to $10^{33}$ years, the Kamiokande experiment was the first to observe supernova neutrinos in 1986, starting a new chapter in the study of neutrino properties. This seems to provide a good example of Weinberg's belief that sometimes science progresses in strange ways through missteps. In his advise to the young scientists, he also said '… forgive yourself for wasting time. ….As you will never be sure which are the right problems to work on, most of the time that you spend in the laboratory or at your desk will be wasted'.

After the discovery of the Cosmic Microwave Background Radiation along with the then newly discovered violation of a symmetry of laws of particle physics, under a combination of charge conjugation and space reflection (CP), Weinberg remarked in the lectures at Brandeis School on Particle Physics that the two discoveries together had the potential to explain why the Universe today is asymmetric in its abundance of matter versus anti-matter, the so-called Baryon Asymmetry of the Universe (BAU). He showed how GUTs can play an important role in explaining the observed BAU. A related work by Weinberg was the systematic development of 'QFT at finite temperature', *finite* in this context meaning non-zero, and at the highest possible energies closest to the Big Bang. It provides the framework for studying the Grand Unification epoch, if there was one, of the Universe. In the sixties Weinberg was the first to appreciate applications of particle physics in astrophysics and cosmology of this variety. Here he was following the early leaders like George Gamow who first thought about CMB and Hans Bethe who elucidated the process of energy generation in the Sun in terms of nuclear reactions.

One of his abiding interests was quantum theory of gravity. He made a most important contribution to the subject through his studies with massless particles, which was almost a matter of obsession with him. He proved that the principles of relativity and quantum mechanics restrict interactions that a massless particle can mediate. He showed that a massless spin-2 boson can couple only to energy-momentum, a property obeyed by gravity and thus making gravity the unique interaction that a massless spin-2 boson can have, consistent with special theory of relativity and quantum mechanics. This observation about massless spin-2 bosons provided a deep understanding of the Equivalence Principle which is the starting point of Einstein's formulation of the General Theory of Relativity (GTR) and indeed opened the doors towards formulation of a quantum theory of gravity, the holy grail of 20th century theoretical physics. His observation meant that a formulation that predicted a massless spin-2 particle, say string theory, could hold the promise of being a theory of quantum gravity.

The need to understand the effects of gravity led him to a thorough investigation of observational cosmology. His scholarly monograph 'Gravitation and cosmology: Principles and applications of the general theory of relativity' published in 1972 heralded the cross-fertilization between the two disciplines of elementary particle physics and cosmology. But it also underpinned a purist philosophy towards theoretical physics. As he put it 'The important thing is to be able to make predictions about the images on the astronomers' photographic plates … and it simply doesn't matter whether we ascribe these predictions to … a curvature of space-time'. As he told Graham Farmelo, he could only be happy as a theorist if experimenters were giving him regular feedback from nature about his speculations.

A very puzzling theoretical issue arising out of GTR, which nettled Einstein himself concerns the possibility of an additional constant term called the 'cosmological constant' in the Einstein equations. Occam's razor of classical reasoning either did not require it at all or required it to be $10^{120}$ times the current energy density of the Universe. Quantum theory insisted it be there and predicted this 'vacuum energy' to have a huge value, smaller than the earlier one but still astronomically large, viz. $10^{55}$ times the current energy density. Such values would have caused the universe to have a very high curvature, in sharp contradiction with the observation of the almost 'flat' and large universe that we find ourselves in. There was some advocacy to settle this scientific issue by suggesting that the small observed value is because it is this value that made human life possible. This is the so called 'Anthropic Principle'. This principle is an attempt to settle something that cannot be scientifically settled. In 1997 Weinberg, with his collaborators, Martel and Shapiro, used this line of thinking and developed an 'Anthropic Criterion' that the observed value of this vacuum energy is what it is because this value then would ensure a Universe where galaxies would be formed and would eventually lead to us being present in the Universe to understand it. Most importantly, using these ideas they gave an order of magnitude estimate of this constant, which was much smaller than the huge value predicted by the quantum theory. The simplest explanation of the astronomical observations of an accelerated expansion of the Universe in 1998 is in fact in terms of a cosmological constant of just about the size computed by Weinberg and collaborators. Weinberg said in an interview to CERN Courier, 'So I was very pleased that the Breakthrough Prize acknowledged some of those things that didn't lead to specific predictions but changed a general framework.

This interest in cosmology also resulted in his famous popular science book, *The First Three Minutes* in 1977, with several more to follow later. Weinberg's touchstone for popular science was that the arguments must remain true to science, yet accessible to an intelligent non-scientist reader such as, understandably in his case, a lawyer. In fact in an interview he had said 'I think it's very important not to write down to the public. You have to keep in mind that you're writing for people who are not mathematically trained but are just as smart as you are.' About his book *Dreams of a Final Theory*, he said in an interview that this future seems further away than we had hoped it to be and that if the promise of string theory as a 'final theory' is borne out, the endeavour of Physics as a 'Mathematical Philosophy of Nature' started by Newton will be finished.

During mid-1980s Weinberg engaged in advocacy for the Superconducting Super Collider (SSC). The W bosons and the Z boson, specific predictions of the EW Theory were discovered at the CERN laboratory in Europe in 1983. A supercollider that would go sufficiently high in energies to also explore supersymmetry was planned to be set up in the USA and was called the SSC. Both SLAC and Fermilab were vying for the project. Texas too made a bid for this accelerator whose ring would be 90 km in circumference and Weinberg was in the committee that prepared the final proposal. The project eventually was awarded to Texas and thus Experimental Particle Physics started at University of Texas at Austin and other campuses in the state. Unfortunately, due to funding issues the project shut down in 1993.

His almost compulsive need to clarify and elucidate led him to pen a series of monographs. Other than the three volumes of QFT and the Gravitation and Cosmology book, he decided in early 2000s to write a modern sequel book *Cosmology*, informed by the latest experimental data and with an aim to arrive at an analytic understanding of the phenomena occurring in the early Universe. Due to the high level of complexity of the concerned equations the process of understanding had been mostly consigned to numerical calculations. The books 'talk' to the reader, avoid scholarly rubric, and yet make a lasting impact as the most enduring scientific monographs. Subsequently his teaching at the undergraduate level resulted in two more books of the textbook genre, accessible to that audience but with the same uncompromising rigour and clarity. These textbooks are an important part of his legacy.

Weinberg considered history of science a useful antidote to philosophy of science. His many articles on science and society, history of science, and science versus organized religion are also published as books. He saw the dangers of rising anti-science sentiment from several lobbies. In 2002 he called upon the students at a college to become his 'allies in a movement … known as the Enlightenment' (The Age of Enlightenment is considered to be approximately the century following Newton's Principia.), because the ethos of the Age of Enlightenment has made the world 'a freer and gentler place' and urged them to guard against the dilution of its values. He encouraged young people to study history of science by saying 'Study the history of science as it will make your work seem more worthwhile to you. Because, a work in science may not yield immediate results, but to realize that it would be a part of history is a wonderful feeling.' He further adds 'As you will learn its rich history, you will come to see how time and time again – from Galileo through Newton and Darwin to Einstein – science has weakened the hold of religious dogmatism'. He felt that the process of research brings one solace and had said in the book *The First*

*Three Minutes*, 'The effort to understand the universe is one of the very few things that lifts human life a little above the level of farce and gives it some of the grace of tragedy'.

During the time one of us (UY) was his research associate, the group would go to the Faculty Club for lunch. Those interactions were very interesting. The group around the table would have members of varied nationalities, Chilean, Spanish, Sri Lankan, Romanian, Swedish, Japanese, British, Belgian and of course Indian. He would usually hold forth on many interesting topics in a voice that would carry and was not deterred by people from other tables telling him how much they enjoyed his 'lecture'.

He had said 'I plan to retire shortly after I die'. Indeed, this prophecy of his too has come to be true. He was teaching and involved in creative research till the end. Among the last publications of his on the iNSPIRE-HEP database website is one from January 2021. It is the text of a talk he gave on EFT, providing guidance to the particle physics community at large. Indeed, this is a perfect example of what he himself had said he always was trying to do: provide a point of view to the fellow physicists. It is humbling and inspiring, that the last entry in this list dated April 2021, is his latest textbook *Foundations of Modern Physics*.

As stated already, UY has had the good fortune to be a graduate student at University of Texas at Austin where Weinberg was on his thesis committee and later took him on as a Postdoctoral Associate. RG, having started her graduate studies in 1974, the year of November revolution which was the first step in establishing the correctness of the SM, is certainly the child of the gauge theory days of particle physics. She has had occasion to listen to his talks/lectures on a number of occasions. In fact, in spite of having 'grown' in SUNY at Stony Brook, the birthplace of Supergravity, the beginning of her journey in Supersymmetry can be traced to lectures on Supersymmetry that Weinberg gave at Austin and a paper by him on Supersymmetry showing that in a certain class of models one would have supersymmetric particles lighter than the W/Z. This latter was one of the early demonstrations of relating this mathematically beautiful and compelling theory to something that could be looked for in experiments going on at the time!! We are very sure that this is just a typical example how Weinberg's work shaped the life of quite a few young particle physicists of the time.

*A note added:* This is a modified version of an obituary to appear in *Physics News*. In this version some details about physics are reduced and more details on the non-physics aspects of his writing and opinion have been added.

Various statements and quotations attributed to Prof. Weinberg and others are taken from one of the following sources:

1. https://home.cern/news/obituary/cern/steven-weinberg-1933-2021
2. https://cerncourier.com/a/still-seeking-solutions/
3. https://cerncourier.com/a/model-physicist/
4. https://grahamfarmelo.com/steven-weinberg-the-universe-speaks-in-numbers-podcast/
5. https://grahamfarmelo.com/remembering-steven-weinberg/
6. https://highprofiles.info/interview/steven-weinberg/
7. https://www.bates.edu/commencement/annual/y2002/honorary-degree-citations/steven-weinbergs-address/
8. https://www.nature.com/articles/426389a
9. https://www.quantamagazine.org/how-steven-weinberg-transformed-physics-and-physicists-20210811/

ROHINI M. GODBOLE[1,*]
URJIT YAJNIK[2]

[1]*Centre for High Energy Physics,
Indian Institute of Science,
Bengaluru 560 012, India*
[2]*Department of Physics,
Indian Institute of Technology,
Mumbai 400 076, India*
*e-mail: rohini@iisc.ac.in

# Kasturi Lal Chopra (1933–2021)

Professor Kasturi Lal Chopra was born to Jagat Ram Chopra and Chanan Devi in Chahal Kalan in Gujranwala district of Punjab on 31 July 1933. He obtained his bachelor's degree in science from the University of Delhi in 1952 with honours and continued his study in the same university to obtain his Master's degree in 1954. Following this, Chopra secured a fellowship at the University of British Columbia and proceeded to North America to carry out his doctoral study. He obtained his Ph.D. from the same university in 1957 working in the field of low-temperature physics. After completing his doctoral study, he joined the Royal Military College of Canada as a Defence Research Fellow to conduct post-doctoral research. After a two-year stay, he moved to Fritz-Haber Institute at Berlin, West Germany, as a Max Planck Fellow from 1959 to 1962. He returned to the USA in 1962 as a research specialist and group leader at Phelico-Ford Scientific Laboratory, Blue Bell, USA. In 1964 he moved to the Ledgemont Laboratory of the Kennecott Copper Corporation as a staff scientist and concurrently Adjunct Professor, North Eastern University, Boston. In 1970, he decided to return to India and joined the Department of Physics at the Indian Institute of Technology, Delhi as a senior Professor of Solid State Physics. He served as department head from 1970 to 1973.

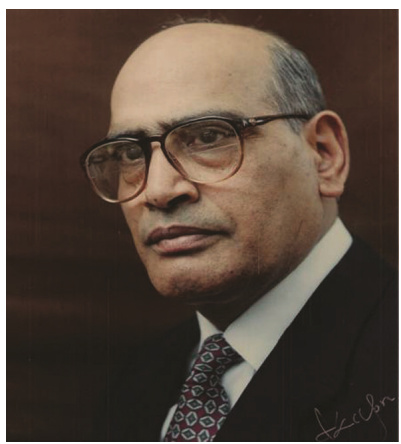

Chopra occupied various academic-related administrative positions at IIT Delhi, as Dean, Faculty of Science (1973–74), as Chairman/Dean, Industrial Research and Development (1975–76 and 1985–87) and Dean, Post Graduate Studies and Research (1976–79). He contributed significantly to the Center for